\documentclass[amsmath,10pt,showpacs,twocolumn,superscriptaddress,aps,pra,floatfix]
{revtex4-1}

\usepackage{graphicx}
\usepackage{multirow}
\usepackage[usenames]{color}
\usepackage{nicefrac}
\usepackage{amsmath}
\usepackage{xparse}
\usepackage{physics}
\usepackage{float}
\usepackage{silence}
\usepackage{xcolor}
\usepackage{hyperref}
\begin{document}

\title{Producing flow in ``racetrack" atom circuits by stirring}
\author{Benjamin Eller}
\author{Olatunde Oladehin}
\author{Daniel Fogarty}
\author{Clayton Heller}
\affiliation{Department of Physics, Georgia Southern University,
Statesboro, GA 30460--8031 USA}
\author{Charles W.\ Clark}
\affiliation{Joint Quantum Institute, National Institute of Standards 
and Technology and the University of Maryland, Gaithersburg, MD 20899, USA}
\author{Mark Edwards} 
\affiliation{Department of Physics, Georgia Southern University,
Statesboro, GA 30460--8031 USA}
\affiliation{Joint Quantum Institute, National Institute of Standards 
and Technology and the University of Maryland, Gaithersburg, MD 20899, USA}
\date{\today}

\begin{abstract}
We present a study of how macroscopic flow can be produced in Bose--Einstein condensates confined in a ``racetrack'' potential by stirring with a wide rectangular barrier.  This potential consists of two half--circle channels separated by straight channels of length $L$ and reduces to a ring potential if $L=0$.  We present the results of a flow--production study where racetrack condensates were stirred with a barrier under varying conditions of barrier height, stir speed, racetrack geometry, and temperature.  The result was that stirring was readily able to produce flow in ring and non--ring geometries but that the exact amount of flow produced depended on all of the study parameters.  We therefore investigated the mechanism by which flow was produced in the stirring process.  The basic mechanism that we discovered was that when the sweeping barrier potential height reached a critical value a series of phase--slip \textcolor{black}{(i.e., a sudden change in the phase winding around the condensate midtrack)} events occurred. Phase slipping stopped when the flow produced overtook the speed of the stirring barrier.  Disturbances generated at each phase slip circulated around the channel and served to convert the initially localized velocity distribution into smooth macroscopic flow.  This picture of the mechanism for making flow should facilitate the design of closed--channel atom circuits for creating a desired amount of quantized smooth flow on--demand.
\end{abstract}


\maketitle
\section{Introduction}
\label{intro}

Recent advances in the optical manipulation of neutral atoms\,\cite{demarco_2008, hadzibabic_2012, boshier_2009, donatella_cassettari} have sparked experimental and theoretical interest in systems of Bose--Einstein--condensed (BEC) atomic gases confined to a thin sheet in a horizontal plane. Cases where the BEC is confined within this plane to a closed--loop channel potential can be roughly analogous to electronic circuits.  The difference is that the current in such ultracold--atom systems refers to the motion of neutral atoms rather than electrons.  These systems are sometimes referred to as ``atom circuits'' and their study is part of the emergent field of atomtronics\,\cite{Amico_2017}.  

Interest in atom circuits derives in part from their potential for use in devices such as rotation sensors\,\cite{Amico_2017} suitable for precision navigation.  Proposed examples include devices that sense rotation via Sagnac interferometry\,\cite{2019arXiv190705466M, 2000Gustavson} and those that act as analogs of Superconducting Quantum Interference Devices (SQUIDs) where rotation takes the place of magnetic flux\,\cite{boshier_2013a, Mathey2016, njp_paper, qs7}. Some implementations of these types of interferometer include a Bose--Einstein--condensed gas confined in a ring geometry\,\cite{Bell2016, ring_circ_probe, sq3, sq7, sq8, 1st_ringBEC_current, 2nd_ringBEC_current, Kumar2016, hysteresis_nature_paper, resistive_flow_BEC}.

\begin{figure}[htb]
\centering
\includegraphics[scale=0.14]{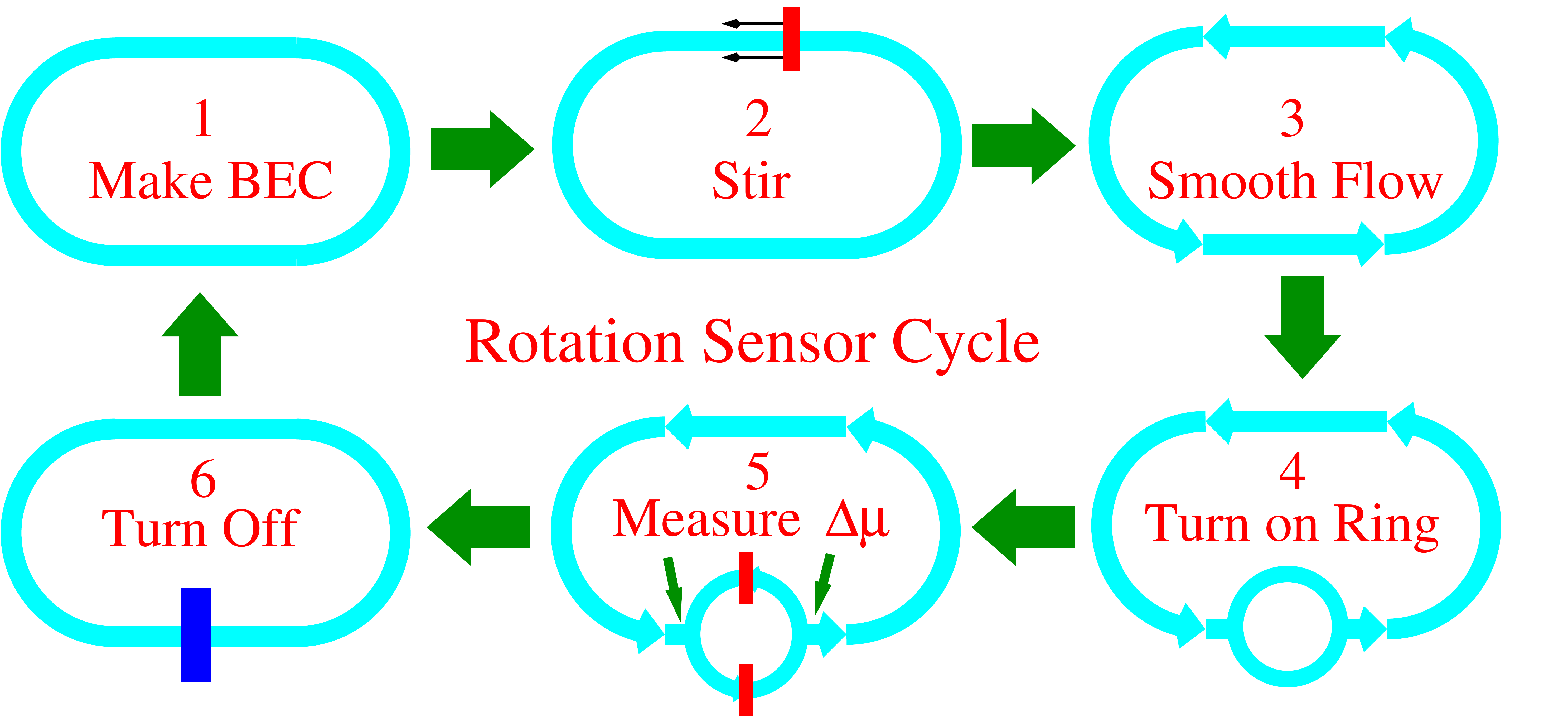}
\caption{Cartoon picture for an imagined cycle of a SQUID--type atom circuit rotation sensor.  \textcolor{black}{The cycle is as follows 1) a condensate is formed in a channel potential, 2) smooth flow is produced by some mechanism, e.g., stirring, 3) the result is a condensate with smooth flow, 4) the potential is then modified to add an inner ring plus Josephson barriers, 5) the difference in the local chemical potential ($\Delta\mu$) is measured, and 6) the system is then reset.}}
\label{squid_sequence}
\end{figure}

All atom circuits require neutral--atom current for their operation. Atom circuits suitable for applications such as rotation sensing mentioned above will need to be able to make \textcolor{black}{repeated} measurements over time.  Such devices will need to step through a cycle where the measurement is made and then reset for the next measurement.  One possible cycle is shown in Fig.\,\ref{squid_sequence} for a SQUID--like rotation sensor\,\cite{boshier_2013a, sackett2014}.  In general terms, the cycle consists of making a BEC, creating flow, modifying the potential to create circuit elements so that the measurement can be made, and then reset so that the cycle can repeat.  Clearly it will be advantageous to be able to create a given amount of smooth flow in the condensate on--demand. 

In order to design these types of atom circuits for applications a detailed understanding of how to produce smooth flow will be essential.  Furthermore, the channel potential that confines the condensate will be modified in each cycle and this will likely require channels that differ from \textcolor{black}{a} ring shape.  This idea motivates our consideration of a ``racetrack'' potential.  The elongated racetrack shape provides extra room for circuit elements to come and go during the cycle.  This potential also has the advantage that the ring shape is a special case.

\textcolor{black}{Finally, if a quantum sensor is to make sensitive measurements, it needs the ability to react to small changes in the environment.  However, these changes may be magnified by the nonlinear behavior of a near--zero temperature condensate (as might happen if the condensate obeyed the Gross--Pitaevksii equation).  If these changes cause large oscillations in the sensor response, its measurement output may be unreliable.  Such oscillations might be controlled by running the sensor at a non--zero temperature.  Thus it is of interest to investigate sensor behavior at non--zero temperature.}

In this paper we investigate the flow--production step of this imagined atom--SQUID sequence by studying how current can be produced in a particular class of atom circuits by stirring.  The atomtronic systems that we will focus on consist of a BEC confined to a horizontal plane in which an arbitrary two--dimensional potential can be created.  

\begin{figure}
\centering
\includegraphics[scale=0.315]{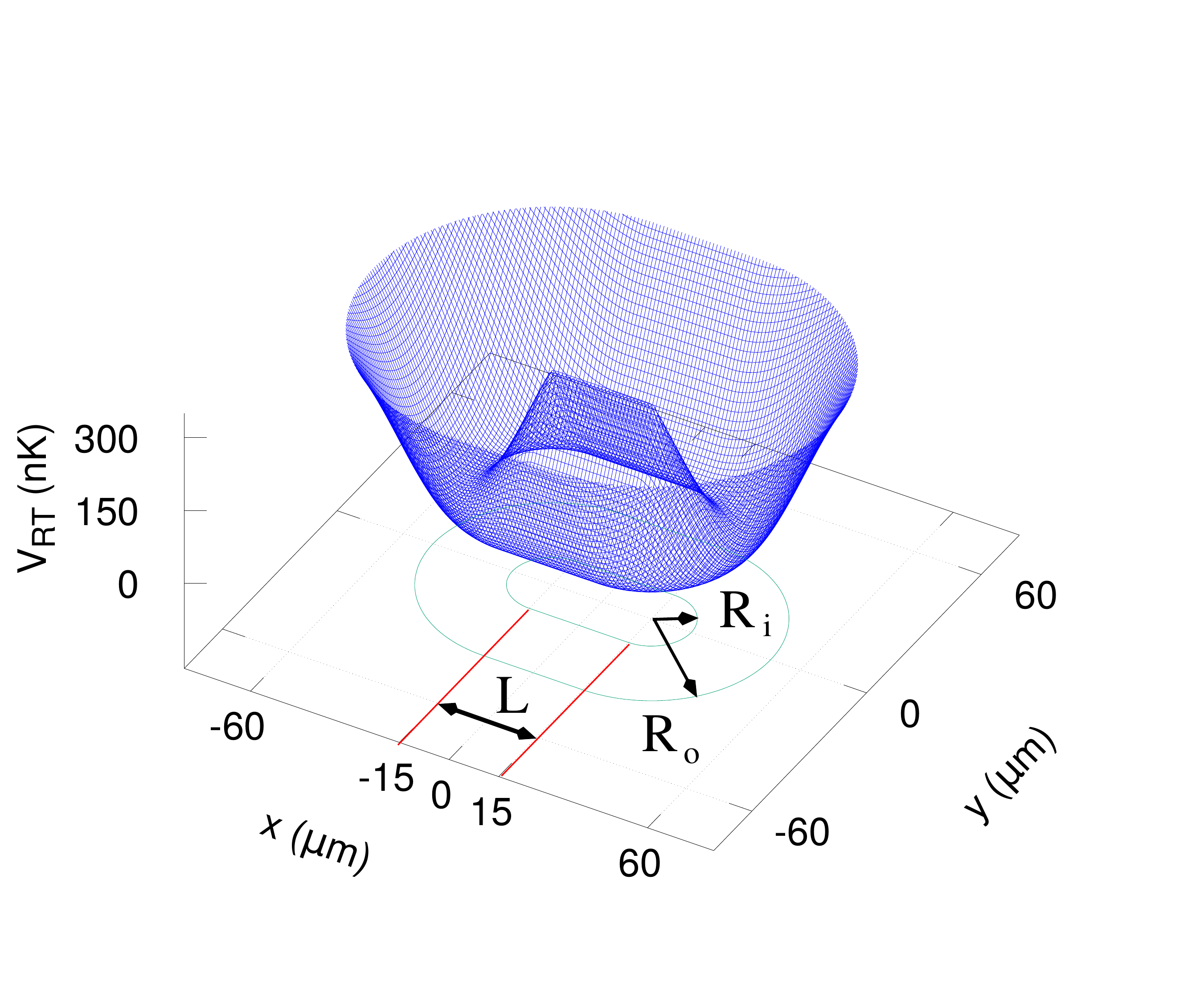}
\caption{A plot of $V_{\rm RT}(x,y)$, which defines the racetrack geometry. The parameter $L$ sets the length of the straight channels that connect the two semi-circular endcaps; depicted here is the $L=30\,\mu$m case. The outer and inner radii parameters, $R_o=36\,\mu$m and $R_i=12\,\mu$m, control the width of the channel. A ring BEC is the $L=0$ special case of the racetrack.}
\label{rt_figa}
\end{figure}

We only considered 2D potentials that take the form of a closed channel in the shape of a racetrack. The racetrack channel consists of two semi--circular endcaps separated by straightaways of length $L$, as illustrated in Fig.\,\ref{rt_figa} and described more fully below.  We also assume that the condensate fills the closed--loop channel entirely.  This differs significantly from some other studies\,\cite{bromley_esry, C_Ryu_2015} where the available volume afforded by the potential was much larger than the size of the condensate so that the potential acts as a waveguide.

Several methods have been used to create flow in BECs confined in ring potentials.  These include transferring orbital angular momentum from a Laguerre--Gauss laser beam to the trapped atoms\,\cite{PhysRevLett.99.260401} and imprinting a phase on the gas atoms using a light pulse with a tailored intensity pattern\,\cite{perrin_phase_imprint}.  The most popular method to--date for producing flow has been stirring the gas with a blue--detuned laser\,\cite{2nd_ringBEC_current, Kumar2016, ring_circ_probe, hysteresis_nature_paper, resistive_flow_BEC, spirals_paper, nist_paper}.  

Here we present a study of the amount of, nature of, and mechanism for creating quantized flow in racetrack BECs by stirring.  In Section\,\ref{ss} we present the results of a systematic flow--production study where racetrack BECs were stirred under different sets of conditions. In these simulations we varied the racetrack lengths, stirring speeds, maximum barrier energy heights, and temperatures.  In Section III we present a detailed account of how stirring produces flow.  In particular we discuss how a single phase--slip\,\textcolor{black}{(i.e., a sudden change in the phase winding around the condensate midtrack)}  occurs and the time sequence of multiple phase slips induced by the stirring. Furthermore we describe how the localized circulation, present just after a phase slip, becomes delocalized macroscopic flow around the ring.  We summarize the results in Section IV.

\section{survey study of flow production by stirring}
\label{ss}

We conducted a survey study of how much flow was produced by stirring a Bose--Einstein condensate, confined in a racetrack channel potential, with a weak--link, rectangular barrier potential. The parameters that were varied in the study were the length, $L$, of the racetrack channel, the thermal--equilibrium temperature, $T$, of the initial state, the stir speed, $v_{b}$ of the barrier, and the maximum energy height of the barrier, $V_{\rm p,max}$. Each simulation in the series was uniquely specified by these parameters: $T$, $L$, $v_{b}$, and $V_{\rm p,max}$.  Except for these parameters, the conditions in all of the simulations were the same.  In this section we describe the full set of conditions present in the simulations and then the flow--production results obtained.

\begin{table*}[htb] 
\caption{Parameter set for the flow--production study. An individual simulation is uniquely identified by the four parameters: $(T,L,v_{b},V_{\rm p,max})$. Each cell containing the label ``$V_{\rm p,max}/\mu=.50,.52,\dots,2.0$'' refers to a unique set of the parameters $(T,L,v_{b})$. This label refers to a set of simulations in which $V_{\rm p,max}/\mu$ varies from $0.50$ up to $2.00$ in steps of $0.02$ where $\mu$ is the chemical potential of the initial condensate.} 
\label{parameters}
\centering
\begin{tabular}[t]{|c || c| c| c| c|}
\hline
$L$\;\;$\rightarrow$ & 
$L=0,30,60\,\mu$m &
$L=0,30,60\,\mu$m &
$L=0,30,60\,\mu$m &
$L=0,30,60\,\mu$m\\
$v_{b}$\;$\downarrow$\;\;$T$\;$\rightarrow$ & 
$T=000$ nK & 
$T=100$ nK & 
$T=150$ nK & 
$T=200$ nK \\
\hline\hline
$113.1\,\mu$m/s & 
$V_{\rm p,max}/\mu=.50,.52,\dots,2.0$ & 
$V_{\rm p,max}/\mu=.50,.52,\dots,2.0$ & 
$V_{\rm p,max}/\mu=.50,.52,\dots,2.0$ & 
$V_{\rm p,max}/\mu=.50,.52,\dots,2.0$\\
\hline
$226.2\,\mu$m/s & 
$V_{\rm p,max}/\mu=.50,.52,\dots,2.0$ & 
$V_{\rm p,max}/\mu=.50,.52,\dots,2.0$ & 
$V_{\rm p,max}/\mu=.50,.52,\dots,2.0$ & 
$V_{\rm p,max}/\mu=.50,.52,\dots,2.0$\\
\hline
$339.3\,\mu$m/s & 
$V_{\rm p,max}/\mu=.50,.52,\dots,2.0$ & 
$V_{\rm p,max}/\mu=.50,.52,\dots,2.0$ & 
$V_{\rm p,max}/\mu=.50,.52,\dots,2.0$ & 
$V_{\rm p,max}/\mu=.50,.52,\dots,2.0$\\
\hline
$452.4\,\mu$m/s & 
$V_{\rm p,max}/\mu=.50,.52,\dots,2.0$ & 
$V_{\rm p,max}/\mu=.50,.52,\dots,2.0$ & 
$V_{\rm p,max}/\mu=.50,.52,\dots,2.0$ & 
$V_{\rm p,max}/\mu=.50,.52,\dots,2.0$\\
\hline
\end{tabular}
\end{table*} 

\subsection{Survey Study Characteristics}
\label{ss_chars}

Here we describe the details of the ultracold--atom system modeled in the simulation, the zero-- and finite--temperature models assumed to govern system behavior, the common characteristics of each simulation, and the ranges of the parameters that were varied.  We begin with the system characteristics.

The initial state of the condensate was assumed to be a stationary thermal--equilibrium system of condensate plus non--condensate held at temperature, $T$. The confining potential present in the initial state was assumed to be strong harmonic confinement in the vertical ($z$ axis) direction plus a ``racetrack'' potential in the horizontal plane.  This potential takes the mathematical form
\begin{equation}
V_{0}({\bf r}) = 
\tfrac{1}{2}M\omega_{z}^{2}z^{2} + 
V_{\rm RT}(x,y),
\end{equation}
where $M$ is the mass of a condensate atom (sodium in this study), $\omega_{z}/2\pi=320$ Hz is the frequency of the vertical harmonic confinement, and $V_{\rm RT}(x,y)$ is the racetrack potential.

The mathematical form for the racetrack potential is given by
\begin{eqnarray}
V_{\rm RT}(x,y) 
&=&  
V_{\rm rt}
\Big\{
\frac{1}{2}\tanh\left(\frac{\rho(x,y)-R_{\rm o}}{\sigma}\right)\nonumber\\
&+&
\frac{1}{2}\tanh\left(\frac{R_{\rm i}-\rho(x,y)}{\sigma}\right)\nonumber\\
&+&
\tanh\left(\frac{R_{\rm o}-R_{\rm i}}{2\sigma}\right)
\Big\} ,
\label{V_rt_eq}
\end{eqnarray}
where the factor $\rho(x,y)$ defines the edges of the condensate in the horizontal plane and is given by
\begin{equation}
\rho(x,y)=\begin{cases} 
      \sqrt{(x-L/2)^2+y^2} & x > L/2 \\
      \sqrt{(x+L/2)^2+y^2} & x < -L/2 \\
      |y| & |x|\leq L/2. 
   \end{cases}
\end{equation}
As illustrated in Fig.\,\ref{rt_figa}, the racetrack potential consists of two half--circular annuli having inner radius, $R_{i}$, outer radius, $R_{o}$, and parallel straightaways of length, $L$. In the simulation study the radii were kept fixed at $R_{i}=12\,\mu{\rm m}$ and $R_{o}=36\,\mu{\rm m}$ while $L$ was one of the parameters that was varied in the simulations.  We chose the racetrack potential because it allows room for adding elements to the atom circuit potential but also enables the well--studied ring case to be recovered for $L=0$.

In each simulation the condensate was stirred by a weak--link rectangular barrier potential, $V_{\rm stir}(x,y,t)$, that swept around the racetrack at constant linear speed, $v_{b}$. The full potential in all simulations had the form
\begin{equation}
V_{\rm ext}({\bf r},t) = 
\tfrac{1}{2}M\omega_{z}^{2}z^{2} + 
V_{\rm RT}(x,y) +
V_{\rm stir}(x,y,t).
\end{equation}
The barrier orientation was always perpendicular to the midtrack and the perpendicular barrier width was always twice that of the channel. Full mathematical details of the racetrack and barrier potentials can be found in Appendix\,\ref{appendixA}.

The height of the barrier was time--dependent.  In all simulations the energy height of the barrier was varied in the same way.  Between times $t=0$ and $t=500$ ms, the barrier energy height was increased linearly from zero to $V_{\rm p,max}$; between times $t=500$ ms and $t=1000$ ms, the energy height was held constant; between times $t=1000$ ms and $t=1500$ ms the barrier was decreased linearly to zero.  For all simulation times $t\ge 1500$ ms, the barrier energy height was zero.  We note that the barrier potential height vs time is plotted in Figs.\,\ref{circ_ring_race}(a) and (b).

In each simulation we allowed the system to evolve for a time after the barrier was turned off.  We did this partly to assess how persistent any flow produced would be and also to be able to implement adding additional elements to the atom circuit.  In this work we only report on the stirring aspect of this sequence.  For the zero--temperature simulations the system was allowed to evolve after the barrier was turned off for a further 2500 ms.  For the non--zero temperature simulations, the system was allowed to evolve only for a further 500 ms.  This reason for this difference was that the non--zero temperature simulation took much more computer time than the zero--temperature ones.  Since the flow--production study required many simulations, for practical reasons we shortened the system evolution time after the barrier was fully off.  

Our choice of barrier stirring protocol is one that has been commonly used in recent experiments\,\cite{2nd_ringBEC_current, Kumar2016, spirals_paper, hysteresis_nature_paper, PhysRevX.4.031052, PhysRevA.95.021602} where flow is produced in ring BECs by stirring.  Our choice was guided by the goal of making {\em smooth} flow and so stirring slowly would minimize unwanted excitations of the condensate.  The stirring speeds in our simulations ranged from $v_{b}=113.1$ to $452.4\,\mu$m/s while bulk sound speeds ranged from $v_{s}=3400$ to $4600\,\mu$m/s. Thus, with our protocol, smooth flow can be produced by stirring at speeds that are a few percent of the bulk sound speed.  Other types of barrier motion such as accelerating barriers where shock waves might be produced\,\cite{njp_paper, Pandey2019} would be less likely to create smooth flow. 

The ranges of parameters $(T,L,v_{b},V_{\rm p,max})$ that were varied in the flow--production study are displayed in Table\,\ref{parameters}.  This set of parameters uniquely identifies an individual simulation.  Each cell of the table labeled ``$V_{\rm p,max}/\mu=.50,.52, \dots,2.0$'' corresponds to specific a value of the temperature, $T$, found at the top of the column, one of the three racetrack length values that also appear at the top of the column, and a specific value of the barrier stirring speed, $v_{b}$, found at the beginning of the row.  The cell label ``$V_{\rm p,max}/\mu=.50,.52, \dots,2.0$'' refers to a series of 76 simulations where the parameters $(T,L,v_{b})$ were the same but $V_{\rm p,max}$ ranged from $0.50\mu$ up to $2.00\mu$ in steps of $0.02\mu$ where $\mu$ is the chemical potential of the initial state.

\begin{figure*}[htb]
\centering
\includegraphics[scale=0.35]{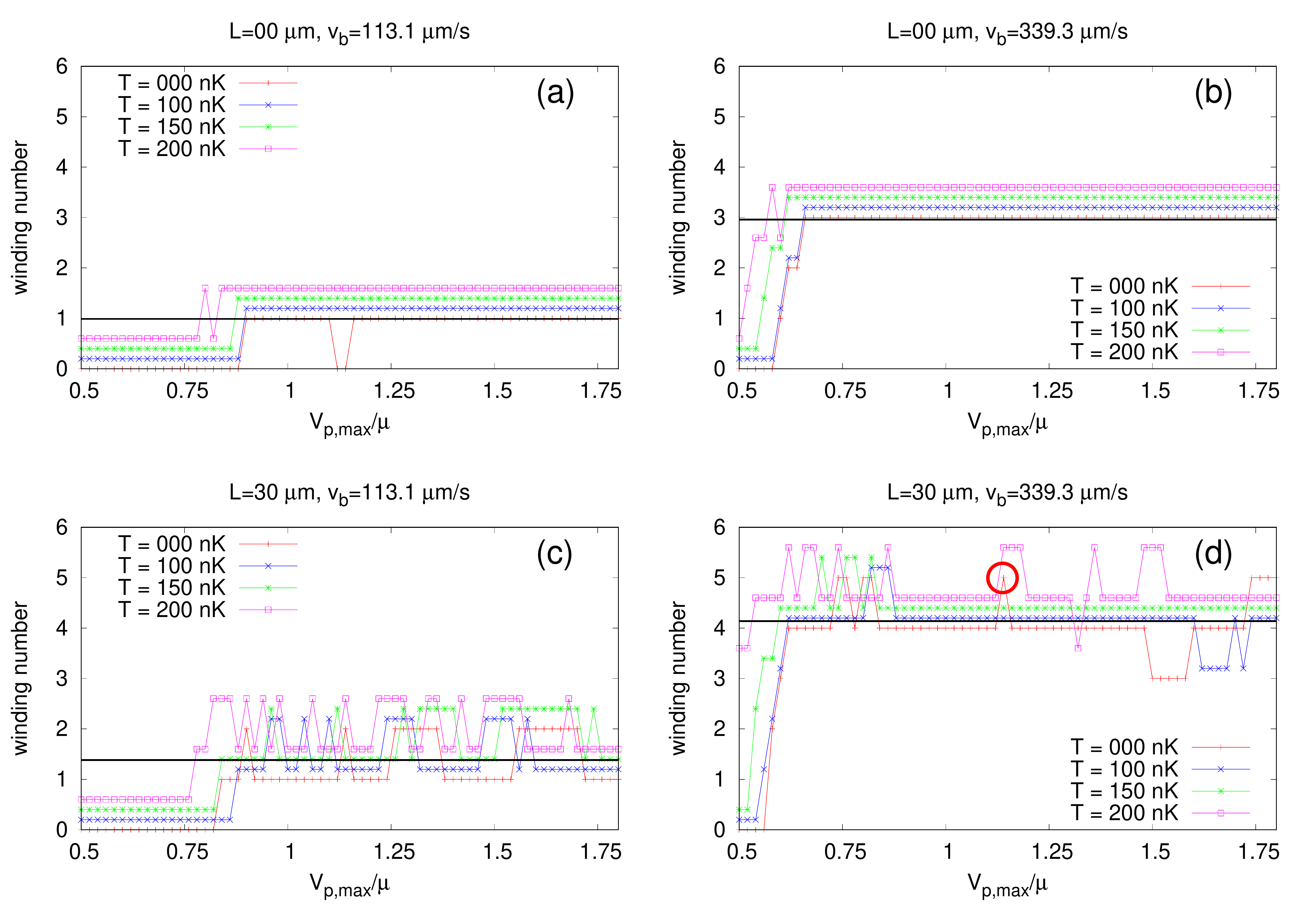}
\caption{Flow produced versus $V_{\rm p,max}/\mu$ in two different racetrack geometries stirring at two different stirring speeds.  (a) $L=0\,\mu$m, $v_{b}=113.1\,\mu$m/s, (b) $L=0\,\mu$m, $v_{b}=339.3\,\mu$m/s, (c) $L=30\,\mu$m, $v_{b}=113.1\,\mu$m/s, and (d) $L=30\,\mu$m, $v_{b}=339.3\,\mu$m/s. Each panel shows the flow produced at four different temperatures: $T = 0, 100, 150,{\rm\ and\ }200$ nK. {\bf Note: The four different temperature curves have been vertically offset for clarity.  The actual value of all flows is the largest integer less than or equal to the values indicated on the curve.} The solid black line indicates the stir speed in units of the flow speed around the midtrack.  The red circle in panel (d) identifies the racetrack case displayed later in Fig.\,\ref{circ_ring_race}(b)}
\label{flow_vs_Vpmax}
\end{figure*}

The behavior of the condensate in zero--temperture simulations was assumed to follow the Gross--Pitaevskii equation (GPE)\,\cite{gross, pitaevskii, pethick_smith_2008}.  For non--zero temperature simulations we used the Zaremba--Nikuni--Griffin (ZNG) model\,\cite{ZNG_yellow_book}.  

In the ZNG model the system is assumed to have a condensate and a non--condensate.  The behavior of the condensate is described by a condensate wave function, $\Phi({\bf r},t)$, and the non--condensate is assumed to be an interacting gas described by a single--particle distribution function, $f({\bf p},{\bf r},t)$.  

The single--particle distribution function is defined so that $f({\bf p},{\bf r},t)d^{3}r\,d^{3}p/(2\pi\hbar) ^{3}$ is the number of particles at time $t$ having position, ${\bf r}$, and momentum, ${\bf p}$.  The non--condensate density, $\tilde{n}({\bf r},t)$ can thus be calculated as
\begin{equation}
\tilde{n}({\bf r},t) = \int\,\frac{d^{3}p}{(2\pi\hbar)^{3}}f({\bf p},{\bf r},t).
\label{noncon}
\end{equation}

The condensate wave function follows a generalized Gross--Pitaevskii equation (GGPE)\,\cite{ZNG_yellow_book}
\begin{equation}
i\hbar\frac{\partial}{\partial t}\Phi({\bf r},t) = 
\Big(
\hat{H}_0 + 2g\Tilde{n}({\bf r},t) -iR({\bf r},t)
\Big)
\Phi({\bf r},t).
\label{GGPE}
\end{equation}
The term $\hat{H}_0=\frac{-\hbar^2}{2M}\nabla^2+ V_{\rm ext}({\bf r},t)+ gn_{c}({\bf r},t)$ is the GPE Hamiltonian, $g$ defines the strength of condensate atom--atom interactions, $n_c({\bf r},t)=|\Phi({\bf r},t)|^2$ is the condensate density, $\tilde{n}({\bf r},t)$ is the non--condensate density and $R({\bf r},t)$ is a local source/sink term that describes particle exchange between condensate and non--condensate.

The single-particle distribution function evolves according to a quantum Boltzmann equation (QBE)
\begin{equation}
\frac{\partial f}{\partial t} - \pmb{\nabla}_{\bf r}U_{\rm eff}\cdot\pmb{\nabla}_{\bf p}f + \frac{\bf p}{M}\cdot\pmb{\nabla}_{\bf r}f = C_{12}[f,\Phi]+C_{22}[f],
\label{QBE}
\end{equation}
where $U_{\rm eff}({\bf r},t)=V_{\rm trap}({\bf r},t) + 2g(n_c({\bf r},t)+\tilde{n}({\bf r},t))$ is an effective potential felt by the non-condensate atoms. \textcolor{black}{The $C_{12}({\bf p},{\bf r},t)$ term is roughly the rate of collisions between condensate and non--condensate atoms with momentum ${\bf p}$ at position ${\bf r}$ and time $t$. These collisions can lead to gain or loss of atoms in the condensate.  The $C_{22}({\bf p},{\bf r},t)$ term describes collisions between two non--condensate atoms at $({\bf p},{\bf r},t)$.  Together these terms describe how collisions affect the rate of change of $f({\bf p},{\bf r},t)$\,\cite{ZNG_yellow_book}.}

The ZNG model allows for the non--condensate density to influence the condensate dynamics at the mean--field level and for the non--condensate density dynamics to couple to the condensate density.  Additionally it allows for particle exchange between condensate and non--condensate via collisions.  

In our simulations we neglected the effect of collisions.  We made this approximation because we believe that the effects of collisions occurs on a much slower timescale than the stirring time (1.5 seconds). The major effect of collisions in our system is the decay of persistent flow \textcolor{black}{which we assume would result from gain or loss of atoms from the condensate. We can put an upper bound of the size of $C_{12}({\bf p}, {\bf r},t)$ ($C_{22}$ should be much smaller) appearing in Eq.\,(\ref{QBE}) by estimating the total possible gain/loss of atoms from the condensate per second due to collisions between condensate and non--condensate atoms.  In the worst--case scenario where $T=200$\,nK so that there are $N_{c}=$\,200,000 condensate atoms and $N_{nc}=$\,300,000 non--condensate atoms there can be about $\approx$\, 12,000 condensate atoms lost per second.  Thus only 18,000 atoms or about 9 percent of the condensate would be lost in a 1.5--second stirring process.  Details of the estimate are found in the footnote.}\,\footnote{\textcolor{black}{This estimate assumes that each non--condensate atom has $n_{c}\sigma\bar{v}\approx 40$ condensate--atom collisions per second. Here $n_{c}$ is the maximum condensate density, $\sigma=8\pi a_{s}^{2}$ is the scattering cross section for atoms and $a_{s}$ is the $s$--wave scattering length and $\bar{v}$ is the average speed of atoms in a gas at $T=200$\,nK. The number of non--condensate atoms available to collide with condensate atoms is estimated as the average non--condensate density in the neighborhood of the condensate times the condensate volume, $N_{cl}=n_{nc}V_{c}\approx 3000$ atoms.  At equilibrium we found the non--condensate density to be about one percent of the maximum condensate density.  Finally we estimate that there is 10\% chance that a collision results in a lost atom\,\cite{pethick_smith_2008, ZNG_yellow_book}}} 

There is also experimental evidence that collisions can be neglected here. At near--zero temperatures, persistent currents induced in ring BECs have a measured lifetime of at least 40 s\,\cite{1st_ringBEC_current} and experimental studies on persistent current decay at non--zero exhibit lifetimes much longer than the 1.5--s stir time\,\cite{PhysRevA.95.021602}.

The ZNG model for dynamics works best in the middle of the range $0 < T < T_{c}$. It has been successfully applied to the damping of collective excitations\,\cite{Jackson_2003} and the decay of an off--center vortex in a simply connected condensate\,\cite{PhysRevA.79.053615}.  It describes both mean--field--dominated regimes and hydrodynamic regimes, except at very low temperatures or in the case of large fluctuations.\,\cite{ Proukakis_2008, zng_chapter}.  The temperatures used in our simulations were chosen by using the ZNG model to compute the condensate fraction versus temperature for a fixed total number of atoms and for the three racetrack lengths as described in Ref.\,\cite{ZNG_yellow_book}.  These curves were fitted using the function:
\begin{equation}
\frac{N_{c}}{N} = 1 - \left(\frac{T}{T_{c}}\right)^{a},
\label{cf_fit}
\end{equation}
with $T_{c}$ and $a$ as fitting parameters\,\cite{PhysRevA.35.4354} and used to select temperatures so that the condensate fractions covered a reasonable range and to ensure the validity of the ZNG model.  These fits yielded critical temperature of $T_{c}\approx 250$ nK.  Thus, at the chosen temperatures, $T=100,150,200$ nK correspond to  $T/T_{c}\approx 0.4,0.6,0.8$, respectively.  Details of how the initial states were calculated along with plots of the condensate fractions vs $T$ can be found in Appendix\,\ref{zng_initial_states}.

\subsection{Survey Study Results}
\label{ss_results}

Typical results of flow production in the BEC by stirring are shown in Fig.\,\ref{flow_vs_Vpmax}.  This figure shows plots of the flow winding number, $n_{w}$, at the end of the simulation versus the maximum energy height of the barrier expressed in units of the chemical potential, $\mu$, of the initial condensate. The winding number is found by computing the phase accumulated around a path along the midtrack of the channel.  Due to the single--valuedness of the condensate wave function, this accumulated phase must equal an integer multiple of $2\pi$ and this multiple is the winding number.

Figure \ref{flow_vs_Vpmax} contains four plots.  Each plot refers to specific values of the racetrack length, $L$ and barrier stirring speed, $v_{b}$.  Appearing in each plot are four curves showing the winding number versus $V_{\rm p,max}/\mu$, one for each of the four temperatures ($T=0,100,150,200$ nK) considered in the flow--production study.  Note that these four curves have been vertically offset for clarity and all winding--number results are integer values.  Each plot also shows a solid (no symbols) black line indicating the stirring speed of the barrier in units of the winding number equivalent to the speed of the stirring barrier.

\begin{figure}[tb]
\centering
\includegraphics[scale=0.65]{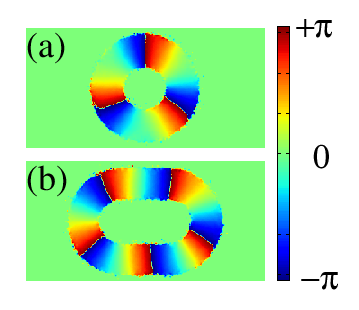}
\caption{Condensate phase distribution in the final state for two stirring cases illustrating the smooth flow obtained by stirring. The rectangles measure $150\,\mu$m horizontally and $75\,\mu$m vertically. (a) Ring case: parameters are $L=0\,\mu$m, $T=0$\,nK, $v_{b}=339.3\,\mu$m/s, $V_{\rm p,max}/\mu=0.98$. (b) Racetrack case: parameters are $L=30\,\mu$m, $T=0$\,nK, $v_{b}=339.3\,\mu$m/s, $V_{\rm p,max}/\mu=1.14$.}
\label{smooth_flow}
\end{figure}

There are several features that are common to all four plots.  First, it is clearly possible to make flow by stirring.  The second notable feature is that no flow is produced until $V_{\rm p,max}$ reaches a critical value and flow is almost always produced for values of $V_{\rm p,max}$ larger than the critical value.  We note that this critical value decreases for increasing initial--state temperature.  This is probably because the total number of atoms in the system is held fixed causing condensate numbers to decrease as $T$ increases.  Finally we see that above the critical value of $V_{\rm p,max}$ the winding number rises rapidly to a plateau after which it oscillates around an average value that is close to the barrier stir speed.  

This average value can be estimated by determining the number of units of flow speed needed to reach the speed of the stirrer.  One unit of average flow speed can be approximated as $\hbar/M$ times the phase gradient around the channel midtrack:
\begin{equation}
v_{\rm flow}=
\frac{\hbar}{M}
\left(
\frac{2\pi}{2\pi R + 2L}
\right)
\label{vflow}
\end{equation}
where $R=(R_{i}+R_{o})/2$ is the average radius of the racetrack endcaps. The stir speed in units of the flow speed, $v_{b}/v_{flow}$ appears as the solid black line in Figs.\,\ref{flow_vs_Vpmax} (a)--(d). This ratio provides a rough estimate of the amount of flow that can be produced by stirring.

\begin{figure*}[tb]
\centering 
\includegraphics[scale=0.30]{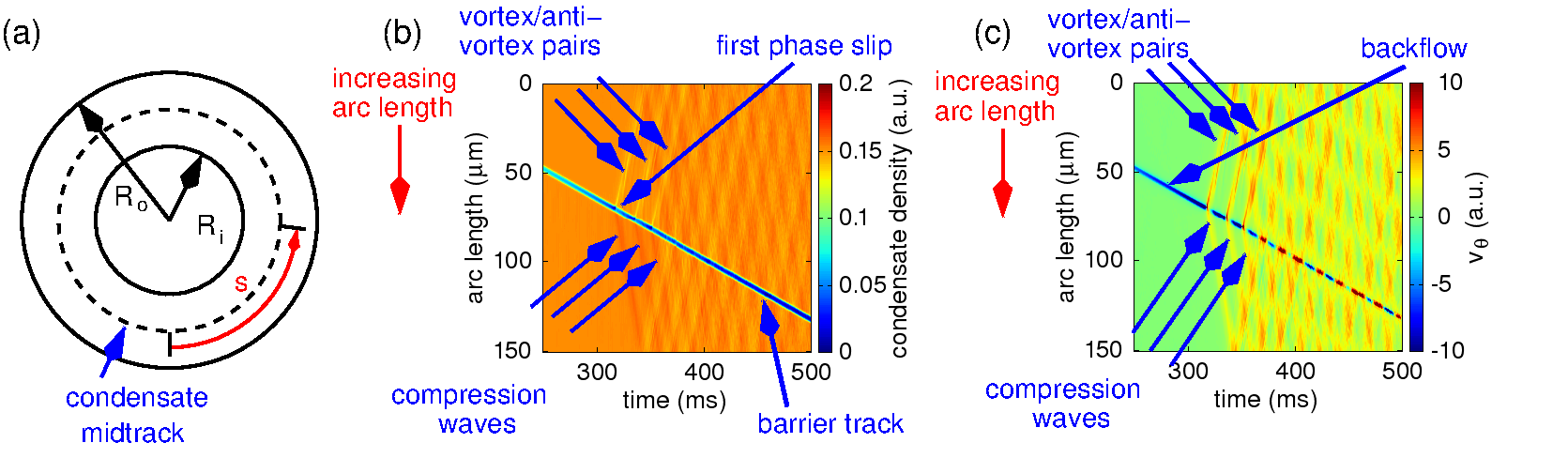}
\caption{Spacetime map of the density and velocity tangential component along the the midtrack of the condensate. (a) The arc length, $s$, is measured from the bottom of the ring and increases in the counterclockwise (ccw) direction.  The barrier also stirs the condensate in the ccw direction. (b) Spacetime map of the condensate density distribution around the midtrack versus time.  The horizontal axis is time and the vertical axis is arc length, $s$, along the midtrack as shown in (a).   (c) Spacetime map of the component of the condensate velocity tangent to the midtrack of the condensate. {\bf Note that the arc length, $s$, increases from top to bottom in panels (b) and (c).}}
\label{1st_phase_slip}
\end{figure*}

Another important question is whether the flow produced by the stirring protocol we have considered here is smooth.  The velocity distribution of the condensate is proportional to the gradient of the phase of the condensate wave function. The signature of {\em smooth flow} along a particular direction is that this spatial rate of change of the phase should be nearly constant.  We can get an indication of whether the flow induced along the channel is smooth by plotting the spatial phase distribution. 

Figure\,\ref{smooth_flow} displays typical final--state phase distributions for a ring and a non--ring case. If we follow the circular midtrack of the ring, see Fig.\,\ref{smooth_flow}(a), we find that the accumulated phase around this path is $3\times 2\pi$ and each $2\pi$ winding takes up very nearly 1/3 of the circumference of this path.  The same is true for the racetrack case, Fig.\,\ref{smooth_flow}(b), where the phase winding divides the midtrack circumference into five approximately equal parts.  From this we infer that the final flow is reasonably smooth.

The full story of the amount of flow produced is more complicated and depends on the details of the time dependence of the barrier turn--on and the shape of the racetrack.  These things can be understood by studying the mechanism of how stirring produces flow within the Gross--Pitaevskii model.  We discuss this in the next section.

\section{How stirring produces flow}
\label{mechanism}

Here we describe how stirring the condensate with a constant--speed barrier whose energy height is increasing produces smooth flow within the Gross--Pitaevskii model.  The basic process is that when the energy height of the barrier exceeds a critical value it triggers a series of phase--slip events causing the accumulated phase around the closed--loop channel to increase. Phase slipping stops when the number of slips times the unit of quantized velocity for the channel, $v_{\rm flow}$, is closest to the stirring speed of the barrier (see the solid black lines in the plots in Fig.\,\ref{flow_vs_Vpmax}).  We note that the details of when and how vortices form at a phase slip has been well--studied\,\cite{Yak1, Yak2, 1st_ringBEC_current, 2nd_ringBEC_current}.  Here we are more concerned with the aftermath of the phase slip and how it contributes to the final macroscopic flow produced.

When the phase--slips stop the tangential component of the condensate velocity is unevenly distributed around the track. This component is large near the vortices created during phase--slip events and small elsewhere. This uneven distribution of velocity is converted into even, smooth flow around the channel during the stirring by pairs of counter--circulating disturbances where each pair is generated at a phase slip.

Hereafter we present the evidence for this narrative of how flow is produced by stirring. We begin by considering how flow is produced in the ring--channel case.  First we describe what happens in a phase--slip event including the nature of the two disturbances generated. Next we present the time sequence of phase slips during the full stirring process.  We also show that the counter--circulating disturbances smooth out fluctuations in the condensate velocity around the channel during the stirring. Finally we return to the racetrack case and describe the effects of a non--ring geometry.

\subsection{Single phase--slip events}
\label{one_flow_unit}

A single phase slip consists of three steps: (1) vortex formation in the barrier near the outer edge of the channel due to condensate backflow inside the barrier region, (2) a vortex/antivortex swap, and (3) generation of two disturbances: a vortex/antivortex pair moving in the anti--stir direction and a compression wave moving in the stir direction. Both disturbances move at the average speed of sound which is much larger than the stir speed of the barrier.  In what follows we shall take the term ``vortex'' to mean a general vortex that circulations in the same direction as the stir and ``antivortex'' to mean one that circulates the opposite way.

\begin{figure*}[tb]
\centering 
\includegraphics[scale=0.28]{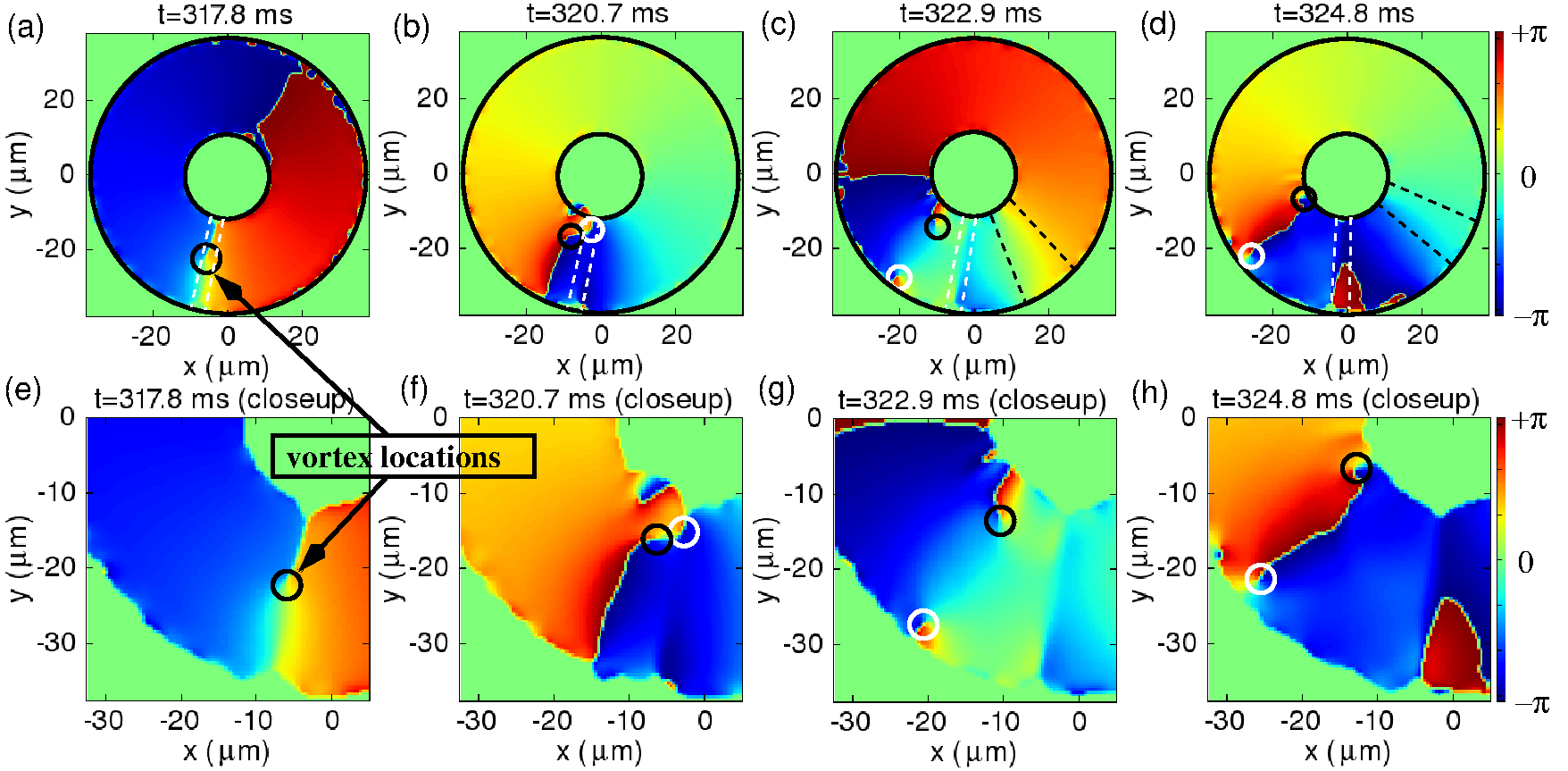}
\caption{Phase distributions for the ring condensate ($L=0\,\mu{\rm m}$, $v_b=339.3 \,\mu{\rm m/s}$, $V_{\rm p,max}=56.9\ {\rm nK}$ and $T=0$ nK, same as in Fig.\,\ref{circ_ring_race}(a)) at times just before and after the first phase slip in Fig.\,\ref{circ_ring_race}(a). The top row of panels, (a)--(d), show the full condensate while panels the bottom row, (e)--(h), show a closeup of the lower--left quadrant of the of the panel just above it.  The large black circles appearing in the top row demark the condensate edges.  Small circles indicate the locations of vortices.  \textcolor{black}{The dotted white lines in the upper row show the approximate position of the stirring barrier.  The wedges marked off by dotted black lines show the approximate position of the compression wave.} Vortices that circulate in the same direction as the stirring (i.e., counterclockwise) are drawn in black. White circles indicate anti--stir circulation.  The times that appear at the top of each picture indicate the time elapsed since the beginning of the stir.}
\label{vd_ring}
\end{figure*}

These steps are illustrated in Figs.\,\ref{1st_phase_slip} and \ref{vd_ring}.  Figures \ref{1st_phase_slip}(b) and (c) show the spacetime distribution of the condensate density, $\rho(s,t)$, at points around the midtrack of the ring and the tangential component of the condensate velocity, $v_{\theta}(s,t)$, around the midtrack, respectively. The horizontal axis is the time, $t$, elapsed since the beginning of stirring and the vertical axis is the arc length, $s$, along the midtrack.  The value of the quantity plotted, $\rho$ or $v_{\theta}$, is represented at each point, $(t,s)$, with a color that can be found in the color bar at the right. As shown in Fig.\,\ref{1st_phase_slip}(a),  the arc length, $s$, increases in the counterclockwise direction as measured from the bottom of the ring. The time interval depicted, $250{\rm\ ms}\le t \le 500{\rm\ ms}$, encompasses the initial series of phase--slips.

The large, mostly blue, stripe running from upper left to lower right and labeled ``barrier track'' in Fig.\,\ref{1st_phase_slip}(b) is the track of the stirring barrier during this time interval.  The stir direction is counterclockwise and so the barrier moves in the positive arc--length direction (top to bottom in the figure).

The barrier stripe also appears in Fig.\,\ref{1st_phase_slip}(c) where the tangential velocity is plotted.  At times before the phase slips begin (labeled by ``backflow'' in the figure) the stripe is deep blue indicating a negative tangential velocity component along the midtrack or backflow in the barrier region. 

When the height of the barrier reaches a critical value, the vortex formed on the outer edge of the channel begins to migrate from the outside to the inside of the channel.  This can be seen in Fig.\,\ref{vd_ring}.  This figure shows a series of phase distribution snapshots during the time interval from just before (see Fig.\,\ref{1st_phase_slip}(b)) until just after the first phase slip.  The color of each point in the plot denotes the value of the phase at that point.  Phase values range from $-\pi$ (blue) up to $+\pi$ (red). Points encircling a vortex core will run through the full spectrum of colors shown in the color bar at the far right of Fig.\,\ref{vd_ring}.  The direction around the circle (clockwise (cw) or counterclockwise (ccw)) going blue to red is the circulation sense of the vortex.

Figures\,\ref{vd_ring}(a) and (e) show the beginning of the migration of the vortex from the outside.  Vortex locations are identified with a circle.  Black circles indicate vortices (i.e., those that circulate in the same sense as the stir) and white circles indicate antivortices.  Panels (a) and (e) show the inward migration of the vortex.  Panels (b) and (f) show the appearance of an antivortex (white circle). Panels (c) and (g) show that the vortex is now on the inner edge and the antivortex is on the outer edge.  The vortex and antivortex ``swap'' places although it is not clear from our simulations exactly where the antivortex comes from.  It is clear that just after the phase slip the vortex and antivortex pair up and move off together in the anti--stir direction.

Shortly after this vortex/antivortex swap, two disturbances are generated.  The first is the vortex/antivortex pair, located on the inside and outside of the channel respectively, that moves away from the barrier in the anti--stir direction.  This can be seen by comparing panels (c)/(g) with panels (d)/(h) of Fig.\,\ref{vd_ring}.  They show that the vortex/antivortex pair has started to move in the anti--stir direction.  This vortex pair causes atoms on the anti--stir side of the barrier to flow in the stir direction.  

The second disturbance is a compression wave that propagates away from barrier region in the stir direction. Evidence for these two disturbances can be seen in Fig.\,\ref{1st_phase_slip}. In panel (b) the annotation ``first phase slip'' points to the location of the barrier when the first phase slip occurs.  Two stripes, a light brown stripe annotated ``vortex/antivortex pair'' and a dark brown stripe annotated ``compression wave'', emanate from the barrier track at the first phase slip point.  

The darker brown color of the compression wave stripe indicates that it is a region of increased density relative to the rest of the condensate. The light brown color of the vortex/antivortex pair stripe shows it to be a region of lower density.  Corresponding stripes for these two disturbances also appear in the tangential velocity plot in panel (c).  We note that both of these are yellow colored indicating that they are both regions of positive (stir direction) tangential velocity while the green regions denote zero tangential velocity.  Thus both disturbances promote condensate flow in the stir direction.  The slopes of the disturbance stripes can be used to determine their speeds. We found that both disturbances move at a speed that is approximately the local speed of sound ($c({\bf r})=\sqrt{gn_{c}({\bf r})/m}$) averaged over the cross section of the condensate.

\subsection{Final flow production: ring case}
\label{ring_final_flow}

Here we describe the overall dynamics of flow production for the ring case. The stirred ring flow dynamics are simpler than for the racetrack and considering the ring case first will enable us to separate effects common to both ring and non--ring cases from those unique to the non--ring geometry.  We will take up the racetrack case in a later section.

\begin{figure*}[tb]
\centering
\includegraphics[scale=0.45]{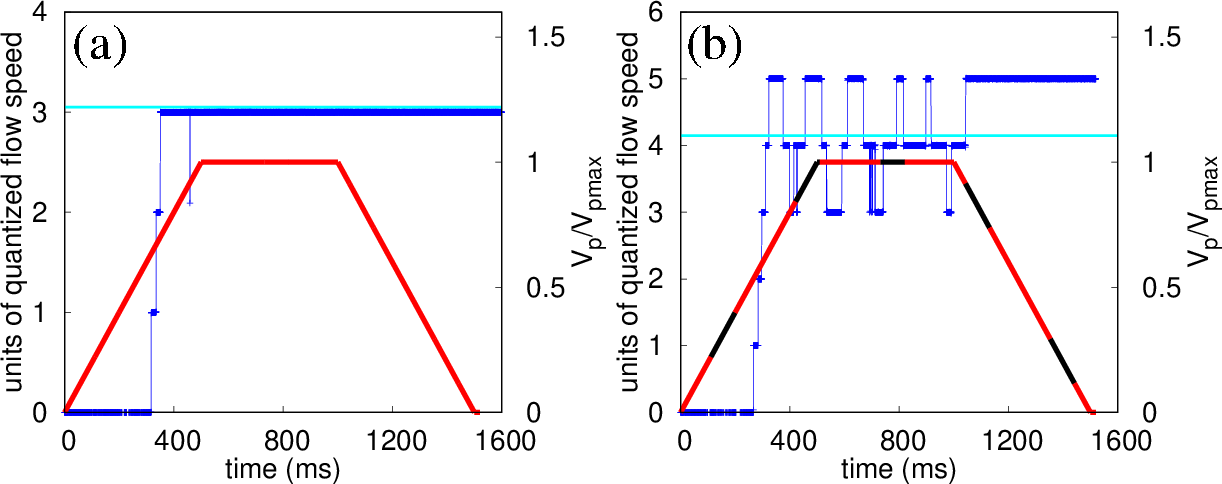}
\caption{(a) Condensate flow speed (blue curve) in units of the quantized flow speed ($v_{\rm flow}=114.6\,\mu{\rm m/s}$ for the $L=0\,\mu{\rm m}$ racetrack) versus time.  The cyan curve shows the stir speed of the barrier in the same units.  The red curve depicts the energy height of the barrier versus time in units of $V_{\rm p,max}$. The case displayed is $L=0\,\mu{\rm m}$, $v_b=339.3\,\mu{\rm m/s}$, $V_{\rm p,max}=56.9\ {\rm nK}$ and $T=0$ nK. (b) Same plot as in (a), except that $L=30\,\mu$m. The black parts of the red--and--black barrier energy--height curve denote times during the stirring when the barrier is on the straightaways.}
\label{circ_ring_race}
\end{figure*}

The typical time sequence for phase slips when the ring condensate is stirred is illustrated in Fig.\,\ref{circ_ring_race}(a) where the blue curve shows the winding number around the midtrack as a function of time during the stirring. The case shown is $L=0\,\mu{\rm m}$, $v_b=339.3\,\mu{\rm m/s}$, $V_{\rm p,max}=56.9\ {\rm nK}$ and $T=0$ nK and is the same case as that depicted in Figs.\,\ref{1st_phase_slip} and \ref{vd_ring}. The vertical axis on the left side of the graph is measured in units of the quantized flow speed, $v_{\rm flow}$.

The red curve indicates the barrier height normalized to its maximum value and the vertical axis on the right side of the graph is the barrier energy height normalized to its maximum value. The cyan curve depicts the speed of the stirring barrier in units of $v_{\rm flow}$.

The behavior of the circulation depicted here is simple: below a critical value of the barrier height, $V_{c}$, there is no circulation, at the critical value three phase slips occur in rapid succession. With each new phase slip, the velocity of the stirring barrier relative to the flowing condensate decreases by one unit of flow speed.  The figure shows that the speed of the flowing condensate overtakes or nearly matches the speed of the barrier.  In this case the backflow that developed when stirring a stationary condensate becomes a {\em forward flow}.  Thus the behavior described earlier that led to the creation of the new units of flow can be reversed and units flow of can be lost.

\textcolor{black}{
We found that the critical barrier height, $V_{c}$, for the onset of phase slips occurs at the same barrier height as long as $V_{\rm p,max} > V_{c}$.  Thus the value of $V_{c}$ (in units of $\mu$) can be inferred from Fig.\,\ref{flow_vs_Vpmax} since $V_{c}$ is the same as the lowest value of $V_{\rm p,max}$ for which maximum flow is obtained.  From Figs.\,\ref{flow_vs_Vpmax}(a) and (c), which show different geometries ($L$) but the same stirring speeds ($v_{b}$), that all of the values of $V_{c}$ lie between 0.75\,$\mu$ and 0.85\,$\mu$.  Comparing Figs.\,\ref{flow_vs_Vpmax}(b) and (d) (again same $L$, different $v_{b}$) we see onset barrier heights lie between 0.55\,$\mu$ and 0.65\,$\mu$.  Thus $V_{c}$ has only a weak dependence on racetrack geometry. }

\textcolor{black}{Comparing onset values from Figs.\,\ref{flow_vs_Vpmax}(a) and (b) (same $L$ different $v_{b}$) we see that faster stirring results in a marked reduction in the onset barrier height $V_{c}$.  The same is true when comparing Figs.\,\ref{flow_vs_Vpmax}(c) and (d).  Finally in each panel of Fig.\,\ref{flow_vs_Vpmax} we can see that $V_{c}$ decreases as the temperature increases.  This may, however, be because we have fixed the total number of particles in the system so that the number of condensate atoms decreases as $T$ increases.
}

\textcolor{black}{
We note that the question of phase--slip production as a function of stirring barrier height has been addressed in the literature\,\cite{Yak1,Yak2}. Our findings for critical barrier height are in line with this previous work.  There has also been previous experimental work on vortex shedding due to a barrier moving through a condensate\,\cite{PhysRevA.91.053615}.  However that work considered a simply connected condensate rather than a multiply connected one. Furthermore, their barrier width was much narrower than their condensate. In contrast our barrier was twice the width of the condensate.  Finally, the barrier speed as a fraction of the bulk sound speed was much higher ($> 30$\%) than in this work where it was less than 10\%.}

\begin{figure*}[tb]
\begin{center}
\includegraphics[width=6.75in]{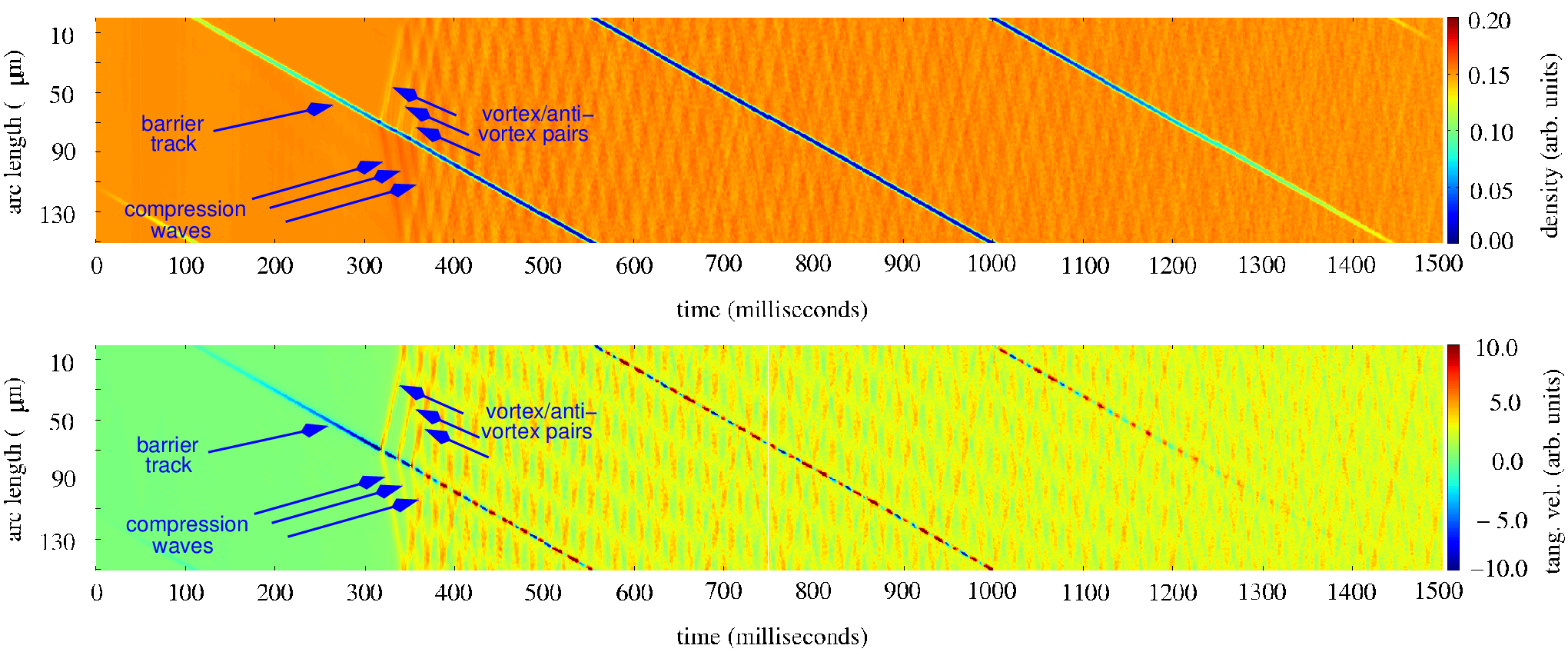}
\end{center}
\caption{Topographic spacetime plots of the density (top) and the tangential velocity component along the channel midtrack (bottom) of the $L=0\,\mu$m (ring) racetrack BEC during the stirring process for the case shown in Fig.\,\ref{circ_ring_race}(a). The large, dark stripe labeled ``barrier track'' appearing in both panels is the track of the stirring barrier. The stripes labeled as ``vortex/antivortex pairs'' show motion in the anti--stir direction while the stripes label as ``compression waves'' show motion in the stir direction. These disturbances convert circulation confined near a localized vortex into into macroscopic flow around the racetrack.}
\label{ring_topo_plots}
\end{figure*}

The three phase slips generate three vortex/antivortex pairs traveling in the anti--stir direction and three compression waves traveling in the stir direction. The behavior of these disturbances during the stirring is depicted in Fig.\,\ref{ring_topo_plots}.  This figure shows the spacetime maps for the condensate density (top panel) and tangential velocity component (bottom panel) for the full duration of the stirring process for the case where $L=0\,\mu{\rm m}$, $v_b=339.3\,\mu{\rm m/s}$, $V_{\rm p,max}=56.9\ {\rm nK}$ and $T=0$ nK.  The same quantities were also shown for a shorter time interval in Figs.\,\ref{1st_phase_slip}(b) and (c).  The phase--slip behavior for this case is shown in Fig.\,\ref{circ_ring_race}(a).

In Fig.\,\ref{ring_topo_plots} both the density and the tangential velocity panel show three pairs of stripes emanating from the barrier track at the times where the series of three phase slips are occurring in Fig.\,\ref{circ_ring_race}(a). The density and tangential velocity panels in Fig.\,\ref{ring_topo_plots} show how these three pairs of disturbances evolve over the duration of the stirring process.  The disturbances continue to circulate around the ring and thereby cause the initially localized velocity distribution to smooth out during the stirring process.  

This is easy to see by looking at the $v_{\theta}(s,t)$ plot in bottom panel of Fig.\,\ref{ring_topo_plots}. Looking at this plot as a whole we can see that it essentially changes color from green to yellow just at the onset of the phase slips at around $t=320$ ms.  However, if we compare this plot for the time interval $350{\rm\ s} \le t \le 600{\rm\ s}$, with the time interval $1250{\rm\ s}\le t \le 1500{\rm\ s}$ at the end of the stirring, we can see that the distribution of velocities around the midtrack is much smoother by the end.

The GPE mechanism for flow production in the ring by stirring with a rectangular barrier can thus be summarized as follows.  The stirring barrier both moves and increases in strength.  This generates a backflow in the region of depressed density in the barrier region.  The backflow causes a vortex to form at the outer edge.  Eventually this vortex migrates inward toward the inner edge of the barrier and a phase slip occurs. This coincides with the appearance of a vortex/antivortex pair with the vortex on the inside and antivortex on the outside. This disturbance moves away from the barrier in the anti--stir direction. At the same time a compression wave disturbance is generated that moves away from the barrier in the stir direction.  These disturbances both move at the average speed of sound.

Phase slips occur in rapid succession until the flow generated overtakes the speed of the stirring barrier.  Each phase slip generates the vortex/antivortex and compression--wave disturbances. These disturbances cycle rapidly around the ring and thereby convert the uneven localized circulation around the ring into evenly distributed flow.  If the generated flow is larger than the barrier speed, the backflow in the barrier region becomes a forward flow and this can cause loss of a unit of flow.  Thus the circulation can oscillate during the stirring period and the final flow amount will depend on how long the stirring period lasts.  We found that oscillations rarely occurred in the ring case.

All of these features are present when flow is created in the non--ring racetrack case.  However, there are some features which only take place for $L\ne 0$ racetrack potentials.  We consider this case next.

\subsection{Final flow production: racetrack case}
\label{racetrack_final_flow}

Many of the features of flow production in the ring condensate are also seen in the racetrack case. In the racetrack case, however, we find features of flow production not present in the ring case. These are (1) that flow oscillations readily occur during the stirring and (2) phase slips seem to take place whenever the barrier moves from curved--to--straight or straight--to--curved parts of the racetrack.  We discuss these new features below.

Comparing the racetrack plot in Fig.\,\ref{circ_ring_race}(b) with the one for the ring in Fig.\,\ref{circ_ring_race}(a) we first see that both plots show a sudden onset of phase slips when the barrier potential reaches a critical value and both show phase slipping continuing until the flow speed overtakes the barrier speed (cyan curve in both plots).  

There is also a striking difference between these two plots: the racetrack plot exhibits oscillations in the flow during the stirring while the ring plot has hardly any.  Most of these oscillations can be understood as the inverse of the phase--slip process described earlier.  Instead of backflow in the barrier causing a vortex to form on the outside, migrating inward and causing a phase slip, a forward flow can develop causing an antivortex can develop on the outside, migrating inward causing a negative phase slip.

\begin{figure}[tb]
\begin{center}
\includegraphics[width=3.0in]{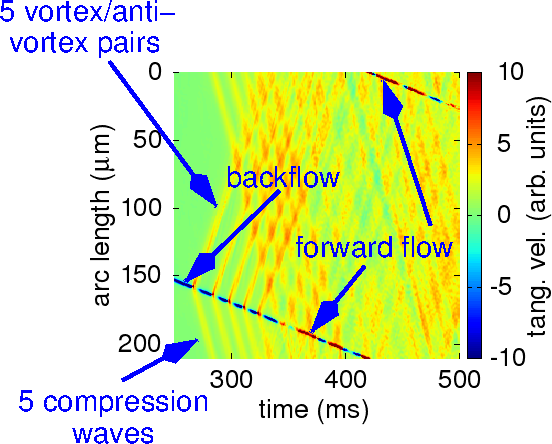}
\end{center}
\caption{Spacetime plot of the tangential component of the condensate velocity along the midtrack for the racetrack case.  the conditions are $L=30\,\mu{\rm m}$, $v_b=339.3\,\mu{\rm m/s}$, $V_{\rm p,max}=56.9\ {\rm nK}$ and $T=0$ nK. \textcolor{black}{The time duration depicted here is during the middle of the stirring sequence.}}
\label{racetrack_phase_slips}
\end{figure}

Figure\,\ref{racetrack_phase_slips} shows how such a forward flow can develop. This annotated figure shows the backflow in the barrier region that causes the phase slips as well as the five pairs of counter--circulating disturbances that are generated. These many disturbances sweep around the racetrack and occasionally intersect each other at the site of the slowly moving barrier.  When this happens the backflow (blue color) in the barrier region can turn into forward flow (red color) since the disturbances generated in the initial phase slips tend to promote flow in the stir direction.  Two sites of such an intersection are shown in the figure and annotated as ``forward flow'' appear at times $t\approx 370$ ms and $t\approx 400$ ms.  These times correlate with flow drops appearing in Fig.\,\ref{circ_ring_race}(b).  Flow can increase because, once the disturbances pass, the backflow reasserts itself and flow--increasing phase slips can occur.  This causes the flow to oscillate during the stir.

Another circulation--changing mechanism that is only present in the non--ring racetrack case occurs when the moving barrier crosses from straight parts of the racetrack channel to curved parts or vice--versa. The times when the barrier is on straight or curved parts are indicated in Fig.\,\ref{circ_ring_race}(b) by the red-- and black--colored curve that depicts the barrier height.  This graph is colored red for times when the barrier is on the curved parts of the racetrack and black--colored when it is on the straightaways. 

Careful examination of the circulation graph shows that, when the barrier transitions from curved to straight (red to black) racetrack parts, the circulation increases by one unit.  When the barrier transitions from straight to curved (black to red) parts the circulation decreases by one unit.  We also note that this only happens when the barrier height is above a certain strength.

It is this mechanism that seems to lead to the final flow value for the case shown in Fig.\,\ref{circ_ring_race}(b).  The very last flow change appearing in this figure is a jump up to five units of flow.  This last jump occurs at $t\approx 1050$ ms just as the barrier moves from the curved end cap to the straightaway as the stirring barrier is beginning to be turned off.  It seems that, by the next transition, the decreasing barrier is too weak to cause any more phase slips.  This particular case differs from other simulations having similar conditions as can be seen by looking at Fig.\,\ref{flow_vs_Vpmax}(d). The racetrack case discussed here is identified there by the red circle.  Note how the final flow for this case is different from those near it on that graph where the only difference is $V_{\rm p,max}$.

It might be possible to avoid this phase slip at the transitions between straight and curved parts of the channel.  Recent work on transport in condensate waveguides,
where the available volume was much larger than the condensate, found that condensates incident on circular bends suffered collective excitations after exiting the bend.  They found that these excitations could be minimized by having a bend in the shape of an ``Euler spiral''\,\cite{C_Ryu_2015, bromley_esry}.  The system we considered here is not a ``waveguide'' in the sense that our condensate occupies the full volume of the channel.  However, modifying the shape of the end caps might eliminate the phase--slip events that occur at the transitions between straight and curved parts of the channel.

\section{Summary}
\label{summary}

We have presented a study of flow production by stirring Bose--Einstein condensates confined in atomtronic racetrack potentials.  We performed a series of simulations under conditions in which the racetrack geometry, initial--state temperature, stir speed, and maximum barrier height were varied.  The study also included an investigation into the mechanism of how flow is produced under the Gross--Pitaevskii model.

We found that stirring is an effective way of creating flow and that there is no difficulty in creating smooth flow in a condensate confined in a non--ring potential.  We also found that flow was readily created when stirring systems initially at finite temperature.

Flow is precipitated by a series of phase slips that appears once the barrier potential reaches a critical height. Each phase slip occurs because a vortex forms on the outside edge of the barrier region due to the buildup of backflow inside the barrier. This vortex migrates to the inner edge of the condensate where it is joined by an antivortex.  

Two disturbances are generated at each phase slip: (1) the vortex/antivortex pair move off in the anti--stir direction and (2) a compression wave that moves in the stir direction. \textcolor{black}{Just after the appearance of the phase slips which generate these disturbances there is a large variation in the distribution of the tangential component of the velocity around the midtrack of the condensate. This can be seen in the bottom panel of Fig.\,\ref{ring_topo_plots}.  } The vortex/antivortex and compression --wave disturbances together promote the formation of smooth flow in the stir direction and facilitate the conversion of localized circulation into macroscopic smooth flow.

The circulation around the racetrack can oscillate because, when one or more of these disturbances simultaneously encounter the barrier region, backflow can be converted into forward flow.  In this case it is possible to have a phase slip in the opposite sense as described above and the total circulation can be decreased by one.

Overall, our results seem to indicate that a user--specified number of units of quantized smooth flow can be generated by stirring on demand.  To make such flow, one only has to match the barrier stirring speed to the number of units of $v_{\rm flow}$ given in Eq.\,(\ref{vflow}).  The value of $v_{\rm flow}$ can be designed by changing the ring or racetrack geometry.  In the non--ring case, one should be careful not be near a curved/straight or straight/curved transition near the time with the barrier height decreases below the critical value for causing a phase slip.

\begin{acknowledgments}
This material is based upon work supported by the National Science Foundation under Grant No.\ PHY--1707776 and by the Physics Frontier Center under Grant No. PHY--1430094. The authors also wish to acknowledge support from the National Institute of Standards and Technology.
\end{acknowledgments}

\appendix

\section{Racetrack and Barrier potentials}
\label{appendixA}

The full potential used in simulating the stirring of a racetrack Bose--Einstein condensate is given by
\begin{equation}
V_{\rm ext}({\bf r},t) = 
\tfrac{1}{2}M\omega_{z}^{2}z^{2} + 
V_{\rm RT}(x,y) + 
V_{\rm stir}(x,y,t).
\end{equation}
The first term represents the vertical harmonic confinement used to restrict the gas to a quasi--two--dimensional horizontal plane. The second term is the racetrack potential that confines the condensate to a racetrack--shaped channel within this plane.  The last term is the potential of the stirring barrier.  We assume that only the first two terms are present for the purposes of defining the initial state.

The racetrack potential is written as a sum of step--up and step--down functions using hyperbolic tangents as follows.
\begin{eqnarray}
V_{\rm RT}(x,y) 
&=&  
V_{\rm rt}
\Big\{
\frac{1}{2}\tanh\left(\tfrac{\rho(x,y)-R_{\rm o}}{\sigma}\right)\\
&+&
\tfrac{1}{2}\tanh\left(\tfrac{R_{\rm i}-\rho(x,y)}{\sigma}\right) +
\tanh\left(\tfrac{R_{\rm o}-R_{\rm i}}{2\sigma}\right)
\Big\} ,\nonumber
\label{V_racetrack}
\end{eqnarray}
where $R_{\rm i}=12\,\mu$m and $R_{\rm o}=36\,\mu$m are the inner and outer radii of the semicircular endcaps.  The factor $\sigma=24\,\mu$m measures the steepness of the step functions. The last hyperbolic tangent term is present above so that the minimum value of the potential is zero.

The factor $\rho(x,y)$ places the jump--up and jump--down sites of the potential thus defining the location of the channel.  It is defined as
\begin{equation}
\rho(x,y)=\begin{cases} 
      \sqrt{(x-L/2)^2+y^2} & x > L/2 \\
      \sqrt{(x+L/2)^2+y^2} & x < -L/2 \\
      |y| & |x|\leq L/2 
   \end{cases}
\end{equation}
where $L$ is the length of the straightaways.

The stir potential is a 2D rectangular barrier whose center coordinates, orientation, and energy height can all have arbitrary time dependence.  The actual potential is most expressed in terms of step--up and step--down functions defined as
\begin{eqnarray}
V_{\rm up}(x,x_{\rm up},\sigma)
&\equiv&
\frac{1}{2}\left[1 + \tanh\left(\frac{x-x_{\rm up}}{\sigma}\right)\right]\nonumber\\
V_{\rm dn}(x,x_{\rm dn},\sigma)
&\equiv&
\frac{1}{2}\left[1 + \tanh\left(\frac{x_{\rm dn}-x}{\sigma}\right)\right]\nonumber
\end{eqnarray}
where $x_{\rm up}$ and $x_{\rm dn}$ denote the places where the step functions equal one--half and $\sigma$ is the steepness of the step.

Using these functions we can write the stir potential as
\begin{eqnarray}
V_{\rm stir}(x,y,t) 
&=&
V_{p}(t)
\Big\{ \nonumber\\
&&
V_{\rm up}(x_{p}(x,y,t),-L_{p}/2,\sigma)\nonumber\\
&\times&
V_{\rm dn}(x_{p}(x,y,t),L_{p}/2,\sigma)\nonumber\\
&\times&
V_{\rm up}(y_{p}(x,y,t),-W_{p}/2,\sigma)\nonumber\\
&\times&
V_{\rm dn}(y_{p}(x,y,t),W_{p}/2,\sigma)
\Big\}
\end{eqnarray}
where $x_{p}$ and $y_{p}$ are barrier coordinates
\begin{eqnarray}
x_p(x,y,x_c(t),y_c(t),\theta_p(t)) 
&=&
(x-x_c(t))\cos(\theta_p(t))\nonumber\\
&+& 
(y-y_c(t))\sin(\theta_p(t))\nonumber\\
y_p(x,y,x_c(t),y_c(t),\theta_p(t)) 
&=&
-(x-x_c(t))\sin(\theta_p(t))\nonumber\\ 
&+& 
(y-y_c(t))\cos(\theta_p(t)).\nonumber
\end{eqnarray} 
Here $x_{c}(t)$ and $y_{c}(t)$ are the time--dependent barrier center coordinates and $\theta_{p}(t)$ is the time--dependent angle that the long dimension of the rectangle makes with the $x$ axis.  The parameters $L_{p}=48\,\mu$m and $W_{p}=3\,\mu$m are the length and width of the barrier, respectively.  The barrier steepness is $\sigma=0.3\,\mu$m.

The barrier center coordinates follow the midtrack of the racetrack and are parameterized using the arc length, $s$, which is measured from the left end of the bottom straightaway:
\begin{equation}
s(t) = s_{0}+v_{b}t \mod{s_{total}}
\end{equation}
where $v_{b}$ is the stir speed, $s_{0}=L+\pi R/2$ is the start point of the barrier stirring, and $s_{total}=2L+2\pi R$ is total arc length of the channel midtrack and where $R=(R_{\rm o}+R_{\rm i})/2$.

\begin{figure*}
\centering
\includegraphics[scale=0.23]{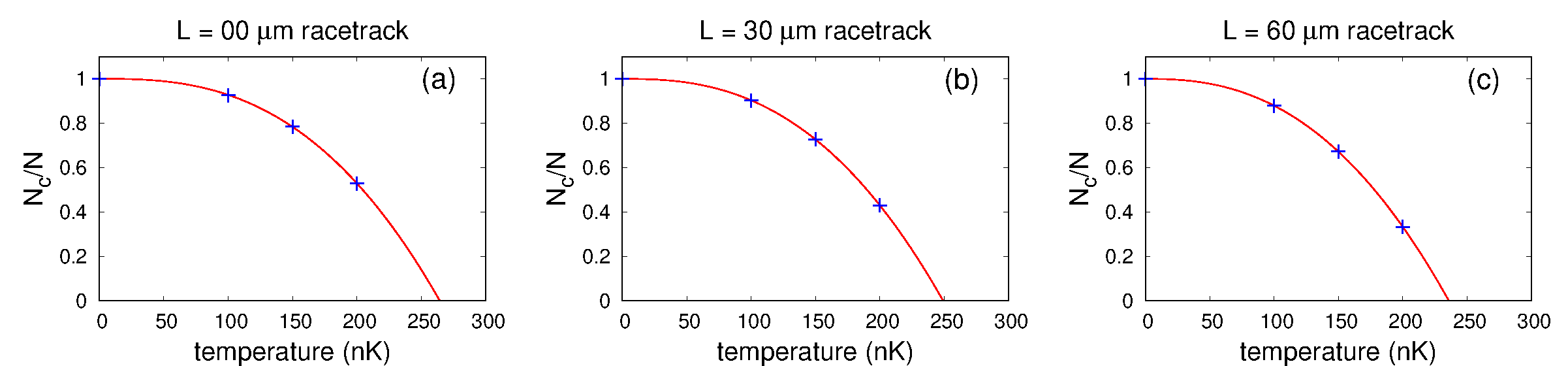}
\caption{ZNG condensate fraction versus temperature for three different racetrack geometries. The blue {\bf +} symbols indicate initial--state condensate fraction as computed by the ZNG model while the solid red line is a fit to the function $N_{c}/N = 1 - \left(T/T_{c}\right)^{\alpha}$. (a) $L=0\,\mu$m, $T_{c}=264.4$ nK, $\alpha=2.697$; (b) $L=30\,\mu$m, $T_{c}=249.3$ nK, $\alpha=2.558$; and (c) $L=60\,\mu$m, $T_{c}=235.8$ nK, $\alpha=2.464$.}
\label{zng_cf}
\end{figure*}

The center coordinates are written in terms of the arc length as
\begin{equation}
x_c(s)=
\begin{cases} 
s-\frac{L}{2} & 0\leq s < L \\
 \frac{L}{2}+R\sin(\frac{s-L}{R}) & L\leq s < s_{1} \\
 \frac{3L}{2}+\pi R-s & s_{1}\leq s < s_{2} \\
-\frac{L}{2}-R\sin(\frac{s-s_{total}+\pi R}{R}) & s_{2}\leq s < s_{total}
\end{cases}\nonumber
\end{equation}
and
\begin{equation}
y_c(s)=\begin{cases} 
      -R & 0\leq s < L \\
      -R\cos(\frac{s-L}{R}) & L\leq s<s_{1} \\
      R & s_{1}\leq s < s_{2}\\
      R\cos(\frac{s-s_{total}+\pi R}{R}) & s_{2}\leq s<s_{total}.
   \end{cases}\nonumber
\end{equation}
Here $s_{1} = s_{total}/2$ and $s_{2}=s_{total}-\pi R$.

The time dependence of the orientation angle is given by
\begin{equation}
\theta_p(s)=\begin{cases} 
      -\frac{\pi}{2} & 0\leq s < L \\
      -\frac{\pi}{2}+\frac{s-L}{R} & L\leq s<s_{1} \\
      \frac{\pi}{2} & s_{1}\leq s<s_{2}\\
      \frac{\pi}{2}+\frac{s-s_{total}+\pi R}{R} & s_{2}\leq s<s_{total}.
   \end{cases}\nonumber
\end{equation}
This dependence orients the barrier so that it is always perpendicular to the midtrack of the channel.

Finally the dependence of the energy height of the barrier on time is written as
\begin{equation}
V_p(t)=\begin{cases} 
      (t/T_{1})V_{\rm pmax}   & 0     \leq t < T_{1} \\
      V_{\rm pmax}            & T_{1} \leq t < T_{2} \\
      (3-t/T_{1})V_{\rm pmax} & T_{2} \leq t < T_{3} \\
      0                       & t     \ge  T_{3}
   \end{cases}\nonumber
\end{equation}
where $T_{1}=500$ ms, $T_{2}=1000$ ms, and $T_{3}= 1500$ ms. This ramps the barrier linearly up to its maximum value, $V_{\rm pmax}$, over a time $T_{1}$, keeps it constant at this value for another time interval $T_{1}$, and ramps it down linearly to zero over yet another time $T_{1}$, and is zero thereafter.

\section{ZNG initial states}
\label{zng_initial_states}

Initial states for the ZNG model are thermal equilibrium states defined by the temperature, $T$, the total number of atoms in the system, $N$, the external potential, $V_{\rm ext}({\bf r})$ (here vertical harmonic plus racetrack), and the atom--atom interaction strength, $g$.  The result of the calculation of the ZNG initial state is a condensate wave function, $\Phi_{0}({\bf r})$, and a non--condensate density, $\tilde{n}_{0}({\bf r})$.  From these, the number of condensate atoms, $N_{c}$ and the chemical potential, $\mu_{0}$ can be obtained.

The iterative method we used to compute these quantities was to start with an initial guess that the non--condensate density was zero, so that $N_{c}=N$, and solve Eq.\,\ref{GGPE} with $R$ and $\tilde{n}$ set to zero.  This yielded a condensate wave function.  This wave function was then used to construct a first guess at the single--particle distribution function.  

In thermal equilibrium, this function has the form\,\cite{ZNG_yellow_book}
\begin{equation}
f^{0}({\bf p},{\bf r}) = 
\frac{1}{e^{\beta_{0}\left[p^{2}/2m+U_{0}({\bf r})-\mu_{0}\right]}-1}
\end{equation}
where, in general,
\begin{equation}
U_{0}({\bf r}) = 
V_{\rm ext}({\bf r}) + 
2g\left[\left|\Phi({\bf r})\right|^{2}+\tilde{n}({\bf r})\right].
\end{equation}
The single--particle distribution function is used to compute a new guess for the non--condensate density using Eq.\,\ref{noncon}. This density is integrated over all position space to obtain a new guess at the number of non--condensate atoms.  This is subtracted from the total number of atoms in the system, $N$, to obtain a new guess at the number of condensate atoms.  This procedure then repeats alternately finding a new condensate wave function and then a new non--condensate density until convergence is achieved.  This procedure is described in more detail in Ref.\,\cite{ZNG_yellow_book}.

This procedure was carried out for the three racetrack geometries $L=0\,\mu$m, $L=30\,\mu$m, and $L=60\,\mu$m for the three different temperatures considered in the survey simulation study, $T=100$ nK, $T=150$ nK, and $T=200$ nK.  The results of these calculations for the condensate fraction versus temperature are shown in Fig.\,\ref{zng_cf}.  The data calculated from the ZNG were fit to the function given in Eq.\,\ref{cf_fit} and these curves are shown in red.

\bibliography{references}
\bibliographystyle{aip2.bst}

\end{document}